\documentclass[superscriptaddress,nofootinbib,twocolumn,eqsecnum]{revtex4-2}

\pdfoutput=1

\usepackage{graphicx}
\usepackage{amssymb,amsmath,latexsym}
\usepackage[colorlinks=true,linktocpage=true,linkcolor=blue,citecolor=orange,urlcolor=blue]{hyperref}
\usepackage{subfigure}

\usepackage{epsfig}
\usepackage{amsfonts}
\usepackage{slashed}
\usepackage[utf8]{inputenc}
\usepackage{epstopdf}
\usepackage{color}
\usepackage[usenames,dvipsnames]{xcolor}
\usepackage{ulem}
\usepackage{comment}
\usepackage{booktabs}

\begin{document}

 \title{Two-flavor chirally imbalanced quark matter beyond large $N_c$}

\author{André G. da Silva }
\email{andre-silva.as@acad.ufsm.br}
\affiliation{Departamento de F\'{\i}sica, Universidade Federal de Santa Maria,
97105-900 Santa Maria, RS, Brazil}

\author{Dyana C. Duarte}
\email{dyana.duarte@ufsm.br}
\affiliation{Departamento de F\'{\i}sica, Universidade Federal de Santa Maria,
97105-900 Santa Maria, RS, Brazil}

\author{Ricardo L. S. Farias}
\email{ricardo.farias@ufsm.br}
\affiliation{Departamento de F\'{\i}sica, Universidade Federal de Santa Maria,
97105-900 Santa Maria, RS, Brazil} 

\author{Marcus Benghi Pinto}
\email{marcus.benghi@ufsc.br}
\affiliation{Departamento de F\'isica, Universidade Federal de Santa Catarina, Florian\'opolis, SC 88040-900, Brazil}
\affiliation{Laboratoire Charles Coulomb (L2C), UMR 5221 CNRS-Universit\'{e} Montpellier, 34095 Montpellier, France}

\author{Rudnei O. Ramos}
\email{rudnei@uerj.br}
\affiliation{Departamento de Física Teórica, Universidade do Estado do Rio de Janeiro, 20550-013 Rio de Janeiro, RJ, Brazil}  

\author{William R. Tavares}
\email{tavares.william@ce.uerj.br}
\affiliation{Departamento de Física Teórica, Universidade do Estado do Rio de Janeiro, 20550-013 Rio de Janeiro, RJ, Brazil}  

 \begin{abstract} 
We investigate a chirally imbalanced medium in the context of the two-flavor Nambu--Jona-Lasinio model using both the large-$N_c$ (LN) and beyond large-$N_c$ (BLN) approximations. To incorporate BLN effects, we consider the optimized perturbation theory (OPT) to the first nontrivial order, which includes two-loop (exchange) contributions. This procedure allows us to explicitly explore how finite $N_c$ corrections affect the thermodynamics as well as the phase diagram of chirally imbalanced quark matter. We then compare the results obtained with a sharp three-dimensional cutoff — generically referred to as the traditional regularization scheme — and with an alternative procedure called the medium separation scheme (MSS). In the first case, we observe that the pseudocritical temperature decreases as the chiral chemical potential increases, an effect dubbed inverse chiral catalysis. On the other hand, when considering the MSS regularization, which properly isolates the medium contributions from the vacuum, we find the opposite result. We show that the results obtained with MSS are consistent with well-established LQCD data in both the LN and BLN approximations. Finally, we suggest that to cope with the high-density limit, the standard OPT interpolation prescription must be modified with the inclusion of an extra variational parameter.
 \end{abstract}

\maketitle

\section{Introduction}
\label{intro}

Effective models like the Nambu--Jona-Lasinio (NJL) model~\cite{Nambu:1961tp,Nambu:1961fr} are invaluable tools for studying quantum chromodynamics (QCD) properties, particularly its thermodynamics, as they capture key nonperturbative features while remaining computationally tractable. The NJL model, for instance, incorporates dynamical chiral symmetry breaking—a crucial aspect of QCD—thus allowing us to explore the phase structure of strongly interacting matter, including the matter formed in the quark-gluon plasma (QGP) at high temperatures and densities. Although it neglects confinement, the simplicity of the model enables analytical and numerical investigations of order parameters, critical phenomena, and the equation of state, providing insights that complement lattice QCD (LQCD) simulations, especially in regimes where lattice methods face challenges, such as at finite chemical potential, a regime in which LQCD is plagued by the sign problem~\cite{deForcrand:2009zkb}. By tuning parameters to fit low-energy QCD observables, the NJL model serves as a phenomenological bridge between theory and experiment, helping us to interpret heavy-ion collision data and astrophysical observations related to neutron stars. Thus, while not a complete substitute for QCD, effective models like the NJL offer a powerful and accessible framework for probing the thermodynamic behavior of strongly interacting matter.

We can also improve on the usual NJL model mean field type of calculations by combining it with alternative nonperturbative resummation techniques, like the optimized perturbation theory (OPT) method~\cite{Okopinska:1987hp,Duncan:1988hw} (see also, e.g., Ref.~\cite{Yukalov:2019nhu} for a recent review and Refs. \cite{Klimenko:1990my,Klimenko:1992av} for seminal papers dedicated to the application of OPT in four-fermion
theories). This combination of different model approximations can provide a powerful framework for improving the study of QCD thermodynamics, particularly in regimes where standard perturbative approaches fail. The OPT method introduces variational parameters that, after being fixed to some optimal value, can capture nonperturbative effects, allowing for better convergence and more reliable predictions at finite temperatures and densities. When applied to the NJL model, it helps to mitigate some of the model's limitations by resumming higher-order contributions in a controlled way. This allows for a more accurate description of the associated phase transitions and critical behavior (for an earlier application of OPT to the NJL model, see, e.g., Ref.~\cite{Kneur:2010yv}). Such a combination is particularly useful for exploring the chiral transition, the equation of state, and the location of the critical end point in the QCD phase diagram, where strong coupling effects dominate. By systematically improving the perturbative expansion, OPT-enhanced NJL models offer a refined tool for theoretical investigations, complementing LQCD and experimental data from heavy-ion collisions.

In general, the nonrenormalizable  $3+1 d$ NJL model (and similar effective models) rely on regularization procedures to handle ultraviolet (UV) divergences. However, when medium effects, such as finite temperature and density, are introduced, improper regularization can lead to nonphysical results. {}For example, usual regularization schemes, e.g., through a sharp cutoff $\Lambda$ in the UV-divergent momentum, which we will designate as the {\it traditional regularization scheme} (TRS), can lead to a mixing of vacuum and medium effects that can distort thermodynamic quantities and break thermodynamic relations, leading to incorrect behavior of the equation of state (EoS) and phase transitions. These issues have recently been addressed through the {\it medium separation scheme} (MSS) regularization procedure~\cite{Farias:2005cr,Farias:2016let,Duarte:2018kfd,Das:2019crc,Lopes:2021tro,Lopes:2025rvn}. The MSS procedure addresses the issues of earlier regularization procedures by systematically separating medium-dependent contributions from vacuum divergences before regularization. The MSS has the advantage of isolating and independently regularizing vacuum and medium parts, ensuring thermodynamic consistency. Hence, this regularization scheme minimizes artificial cutoff effects on thermodynamic observables and better captures the correct asymptotic behavior of QCD-inspired models, making it more reliable for studies of dense matter in astrophysics and heavy-ion collisions.

Physical systems with nonzero chirality have become a subject of considerable attention because of the wide range of anomalous phenomena they can exhibit. Among these is the chiral magnetic effect (CME), where a magnetic field induces a vector current in the presence of a chiral imbalance~\cite{Fukushima:2008xe,Vilenkin:1980fu}; the chiral separation effect (CSE), which refers to the generation of an axial current by a magnetic field in both quark and baryonic matter~\cite{Son:2004tq,Metlitski:2005pr}; and the chiral vortical effect (CVE), whereby rotational motion in a relativistic fluid generates a current~\cite{Vilenkin:1979ui,Banerjee:2008th,Son:2009tf,Landsteiner:2011cp}. The interplay between the CME and external magnetic fields can further enhance these effects. {}Furthermore, the effect of chiral imbalance, combined with electric fields~\cite{Ruggieri:2016xww} and isospin chemical potential~\cite{Khunjua:2018sro,Khunjua:2018jmn,Khunjua:2020vrp,Khunjua:2021oxf}, has also been explored in the context of the chiral phase transition. In addition, chiral imbalance can cause phenomena such as the chirality-induced Kondo effect~\cite{Suenaga:2019jqu} and has been linked to other mechanisms~\cite{Kharzeev:2010gd,Rajagopal:2015roa,Yamamoto:2015ria,Chernodub:2015gxa,Sadofyev:2015hxa}. A comprehensive overview of applications involving chiral-type effects can be found in~\cite{Yang:2020ykt}. These include heavy-ion collisions~\cite{Kharzeev:2024zzm}, studies of Weyl and Dirac semimetals~\cite{Song:2016ufw,Braguta:2019pxt}, the early Universe~\cite{Kamada:2022nyt}, and compact astrophysical objects~\cite{Charbonneau:2009ax,Ohnishi:2014uea}.

To study chiral imbalance in thermal field theories, a chiral chemical potential, denoted $\mu_5$, is introduced in the grand partition function. A notable advantage of introducing $\mu_5$ is that, contrary to the case with finite baryon chemical potential $\mu_B$, first-principle simulations of QCD in the lattice are free from the sign problem. Relatively recent studies~\cite{Braguta:2015zta,Braguta:2015owi} have shown, for instance, that the critical temperature for restoration of chiral symmetry increases with $\mu_5$, in contrast to the model predictions drawn from the NJL~\cite{Ruggieri:2011xc,Fukushima:2010fe,Chao:2013qpa,Yu:2014sla,Yu:2015hym} and linear sigma models~\cite{Ruggieri:2011xc,Chernodub:2011fr}. {}Further support for the behavior of the critical temperature as a function of $\mu_5$, observed within the lattice simulations, can be  found in different approaches. These include large-$N_c$ (LN) universality arguments~\cite{Hanada:2011jb}, Dyson-Schwinger equations with effective quark-gluon interactions~\cite{Wang:2015tia,Xu:2015vna,Shi:2020uyb,Tian:2015rla,Cui:2016zqp}, nonlocal NJL models~\cite{Frasca:2016rsi,Ruggieri:2016ejz,Ruggieri:2016cbq,Ruggieri:2020qtq}, self-consistent mean-field treatment within the NJL model~\cite{Yang:2020ykt,Liu:2020elq,Liu:2020fan,Yang:2019lyn,Pan:2016ecs,Lu:2016uwy}, nonstandard renormalization in the quark linear sigma model~\cite{Ruggieri:2016cbq,Ruggieri:2016ejz}, and frameworks based on chiral perturbation theory~\cite{Espriu:2020dge,GomezNicola:2023ghi,Andrianov:2019fwz}. All these results point toward the argument put forward in Ref.~\cite{Braguta:2016aov}, namely that $\mu_5$ enhances quark--antiquark pairing, which increases the quark condensate and, as a consequence, raises the temperature required for chiral symmetry restoration. This outcome is consistent with the lattice predictions.

The correct description of the local NJL model was first achieved in Ref.~\cite{Farias:2016let}, through the implementation of the MSS. A subsequent study~\cite{Azeredo:2024sqc} employed a new parametrization of the Polyakov loop potential together with MSS, successfully reproducing lattice-consistent results within the PNJL model at a finite chiral chemical potential. In the present work, we apply the MSS regularization procedure, for the first time combined with the OPT method, to study a cold and chirally imbalanced two-flavor NJL model. This system has been of recent interest, particularly because it has been claimed to exhibit inverse chiral catalysis (ICC)~\cite{Ghosh:2023rft,Ghosh:2025zkk} within the TRS type of regularization, which is not consistent with LQCD. Here, we show that new contributions from the beyond-large-$N_c$ (BLN) OPT approximation do not change the qualitative behavior of the pseudocritical temperature as a function of the chiral chemical potential, indicating that the MSS procedure is unavoidable at this point.

In this paper, we advance the NJL model framework beyond previous studies in three key aspects. We start by presenting the first application of the MSS regularization scheme combined with the OPT method. This combination improves upon earlier MSS implementations in the NJL model, which relied solely on the LN limit, by systematically incorporating nonperturbative corrections via the OPT method. Second, by introducing chiral imbalance effects alongside the OPT-MSS framework, we significantly extend the model's applicability to realistic systems, such as dense matter under chiral asymmetry. Third, we demonstrate the need to improve the standard implementation of the OPT method for dense systems, which implies the addition of a novel second variational parameter. This modification of the original prescription (e.g., adopted in Ref.~\cite{Kneur:2010yv}) is fundamental for the proper description of the system at high densities.

The remainder of the paper is organized as follows. In Sec.~\ref{model}, we present the basic structure of the  $SU(2)$ NJL model within both LN and OPT to first nontrivial order, followed by the details of the TRS and MSS regularizations procedures. The numerical results are shown in Sec.~\ref{NR}, while our conclusions are presented in Sec.~\ref{conclusions}.

\section{The chirally imbalanced NJL model}
\label{model}

In order to explore the thermodynamics of the NJL model at finite densities and temperatures, and analyze how its phase structure is influenced by the presence of a chiral imbalance between left- and right-handed quarks, let us consider the following $SU(2)$ Lagrangian density (written in natural units)
\begin{equation}
\mathcal{L} = \bar{\psi}\left(i{\partial \hbox{$\!\!\!/$}}
 - m_c\right)\psi + G\left[\left(\bar{\psi}\psi\right)^2 +
   \left(\bar{\psi}i\gamma_5\vec{\tau}\psi\right)^2\right]\,,
 \label{lagr}    
\end{equation}
where $\psi$ represents a flavor isodoublet for $u,d$ quarks, of $N_c$-plet quark fields, and $\vec{\tau}$ are the Pauli matrices in the isospin basis. In Eq.~(\ref{lagr}), $m_c$ represents the current quark mass ($m_u = m_d$ in the isospin symmetric regime), while $G$ represents the coupling constant for the scalar and pseudoscalar sectors, parametrizing the four-fermion interactions. 

In the LN approximation, which is equivalent to the mean-field approximation~\cite{Kneur:2010yv}, one usually introduces auxiliary bosonic fields $\sigma$ and $\vec{\pi}$ through a Hubbard-Stratonovich transformation. Then, a bosonized version of the model can be written as 
\begin{eqnarray}
\mathcal{L}_{\text{LN}} &=& \bar{\psi}\left(i{\partial \hbox{$\!\!\!/$}}
 - m_c\right)\psi
- \bar{\psi}\left(\sigma + i\gamma_5\tau\cdot\vec{\pi}\right)\psi\nonumber\\
&& 
- G(\sigma^2 + \vec{\pi}^2).
\end{eqnarray}

To implement the OPT within the NJL model, one can adopt the procedure outlined in Refs.~\cite{Kneur:2006ht,Kneur:2007vm,Kneur:2007vj}, which consists of interpolating the original four-fermion interaction using a fictitious bookkeeping expansion parameter, $\delta$. {}Following this approach, the deformed Lagrangian density of the NJL model, expressed in terms of the auxiliary fields, becomes
\begin{eqnarray}
\mathcal{L}_{\text{OPT}} &=& \bar{\psi}\left[i{\partial \hbox{$\!\!\!/$}}
 - m_c - \delta(\sigma + i \gamma_5\vec{\tau}\cdot\vec{\pi}) - (1-\delta)\eta\right]\psi
\nonumber\\
&&- 
\delta G(\sigma^2 + \vec{\pi}^2).
\label{LagOPT}
\end{eqnarray}
Note that the original bosonized Lagrangian density is recovered by setting $\delta = 1$ in the above expression. Here, $\eta$ is an arbitrary mass parameter, which will later be fixed at some optimal value, $\overline \eta$. In Ref.~\cite{Gandhi:1992xd} it was shown that $\eta$ can be extended to account for arbitrary mass parameters in the pseudoscalar direction, through the redefinition $\eta\to\eta + i\gamma_5\tau\cdot\vec{\beta}$ so as to preserve the underlying symmetries. However, from the thermodynamic potential, the physically relevant fluctuations are in the scalar direction, since only the scalar field $\sigma$ develops a nonzero vacuum expectation value. This choice reflects the spontaneous breaking of chiral symmetry along the scalar channel, while the vacuum expectation values of the pseudoscalar fields, $\pi_i$, vanish. Consistently, within the OPT framework, this implies $\vec{\beta}_i = 0$~\cite{Gandhi:1992xd}. More details on the procedure to determine $\eta$ will be provided below.

At finite temperature $T$, quark chemical potential $\mu$, and chiral chemical potential $\mu_5$, the thermodynamics can be studied from the partition function in the grand canonical ensemble,
\begin{align}
\mathcal{Z}&(T,\mu,\mu_5) =
\int[d\bar{\psi}][d\psi]\times\nonumber\\ &\exp\left[\int\limits_0^\beta
  d\tau\int d^3x\left(\mathcal{L} +
  \bar{\psi}\mu\gamma_0\psi +
\bar{\psi}\mu_5\gamma_0\gamma_5\psi\right)\right]\,,
\label{ZTmu}
\end{align}
where $\mu = \text{diag}(\mu_u,\mu_d)$ represents the quark chemical potential, which relates to the baryon chemical potential as $\mu =\mu_B/3$ in the isospin symmetric limit. At the same time, 
$\mu_5$ represents the pseudochemical potential related to the imbalance between left- and right-handed quarks. In the OPT implementation for this model, it is convenient to also introduce an additional variational parameter $\zeta$, through the replacement 
$\mu\to \mu+ (1-\delta)\zeta$,
analogous to the replacement made for the current mass,
$m_c \to m_c + (1-\delta) \eta$ when writing the interpolated Lagrangian density given in Eq.~(\ref{LagOPT}). The inclusion of this additional variational parameter turns out to be fundamental in order to ensure that the optimized results converge correctly at high densities, as we shall further discuss when presenting our numerical results. By employing the corresponding Lagrangian density in the partition function and following Ref.~\cite{Kneur:2010yv}, one can write the free energy density at order $\delta$ as
\begin{eqnarray}
    \mathcal{F}_{\text{OPT}} = &&\frac{(M - m_c)^2}{4G} - N_f N_c I_1(T, \tilde\mu, \mu_5) \nonumber\\
    &&+ \delta N_f N_c (\eta + m_c)(\eta - M + m_c) I_2 (T, \tilde\mu, \mu_5) \nonumber\\
    && + \delta N_f N_c \zeta I_3(T, \tilde\mu, \mu_5) + \delta G N_f N_c I_3^2(T, \tilde\mu, \mu_5) \nonumber\\
        &&- \frac{1}{2} \delta G N_f N_c (\eta + m_c)^2 I_2^2(T, \tilde\mu, \mu_5),
        \label{freeOPT}
\end{eqnarray}
with the definitions 
\begin{eqnarray}
    I_1(T, \tilde\mu, \mu_5) = &&\sum_{s = \pm 1} \int \frac{d^3p}{(2 \pi)^3} \Big[ E_p + T \ln( 1 + e^{-(E_p - \tilde\mu)/T}) \nonumber\\ 
    &&+ T \ln( 1 + e^{-(E_p + \tilde\mu)/T}) \Big],
\end{eqnarray}
\begin{eqnarray}
    I_2(T, \tilde\mu, \mu_5) = &&\sum_{s = \pm 1}  \int \frac{d^3p}{(2 \pi)^3} \frac{1}{E_p} \bigg( 1 - \frac{1}{e^{(E_p - \tilde\mu)/T} + 1} \nonumber\\
    && - \frac{1}{e^{(E_p + \tilde\mu)/T} + 1} \bigg),
\end{eqnarray}
\begin{eqnarray}
    I_3(T, \tilde\mu, \mu_5) = &&\sum_{s = \pm 1} \int \frac{d^3p}{(2 \pi)^3} \bigg(\frac{1}{e^{(E_p - \tilde\mu)/T} + 1} \nonumber\\
    && - \frac{1}{e^{(E_p + \tilde\mu)/T} + 1} \bigg),
\end{eqnarray}
where $\tilde\mu = \mu + \zeta$ and $M = {\overline \sigma} + m_c$ (with $\langle \sigma \rangle \equiv {\overline \sigma}$).
The dispersion relation is
\begin{equation}
    E_p = \sqrt{(|\textbf{p}| + s \mu_5)^2 + (\eta + m_c)^2}.
\end{equation}

It is important to note that even at this first nontrivial order, the OPT introduces finite $N_c$ corrections. This can be easily understood by recalling that, within the LN approximation, the coupling $G$ is of order $N_c^{-1}$, so that the last two terms in Eq.~(\ref{freeOPT}) are of order $N_c^0$, while all others are of order $N_c$. In this context, the density dependent\footnote{Note that $I_3=0$ for $\tilde \mu=0$.} term $\delta G N_f N_c I_3^2$ is of particular importance. As noted in Refs.~\cite {Ferroni:2010ct,Restrepo:2014fna} this two loop contribution is similar to the term $G_V N_c^2 N_f^2 I_3^2 = G_V n^2 $ that appears in mean field evaluations performed within the NJL model when a repulsive channel, parametrized by $G_V$, is considered [see  Ref.~\cite{Klevansky:1992qe} for more details on the effects of exchange (Fock) type of terms].
At the same time, it is a well-established fact that the presence of repulsive vector contributions tends to stiffen the EoS while rendering the phase transitions less abrupt \cite{Fukushima:2008wg,Fukushima:2008is}. Therefore, since here  $G_V=0$, we may expect that the OPT will predict smoother phase transition patterns and stiffer  EoS than the LN approximation. 

{}For completeness, let us also quote the standard LN (or meanfield) result~\cite{Klevansky:1992qe,Buballa:2003qv}
\begin{equation}
    \mathcal{F}_{\text{LN}} = \frac{(M - m_c)^2}{4G} - N_f N_c I_1(T, \mu, \mu_5),
\end{equation}
where $I_1(T,\mu, \mu_5)$ for LN can be obtained from $I_1(T,\tilde{\mu}, \mu_5)$ in OPT expression by replacing
$(\eta + m_c) \rightarrow M$ and $\zeta \rightarrow 0$. 
Since in this work we are mainly concerned with cold and dense matter, it is useful to take the $T\to 0$ limit in all of the thermal integrals above. {}For $I_1$, $I_2$, and $I_3$ this yields the following explicit results depending on the chemical potential,
\begin{eqnarray}
    I_1(&&T = 0, \mu, \mu_5) - I_1(T = 0, \mu = 0, \mu_5) \nonumber\\
    &&= \frac{\theta(\mu - \eta - m_c)}{16\pi^2} \Bigg\{(\eta + m_c)^2 \Big[ (\eta + m_c)^2 - 4\mu_5^2 \Big] \nonumber\\
    &&\times \ln \bigg[ \frac{(\sqrt{\mu^2 - (\eta + m_c)^2} + \mu)^2}{(\eta + m_c)^2} \bigg] \nonumber\\
    &&+ \frac{10}{3} \mu \Big[ \mu^2 - (\eta + m_c)^2 \Big]^{3/2}  - 2\mu^3 \sqrt{\mu^2 - (\eta + m_c)^2} \nonumber\\
    &&+ 8\mu \mu_5^2 \sqrt{\mu^2 - (\eta + m_c)^2} \Bigg\},
\end{eqnarray}
\begin{equation}
\begin{aligned}
    I_2(&T = 0, \mu, \mu_5) - I_2(T = 0, \mu = 0, \mu_5) \\
    &= -\frac{\theta(\mu - \eta - m_c)}{2\pi^2}\Bigg\{ \mu \sqrt{\mu^2 - (\eta + m_c)^2} \\
    &+ \bigg[ \mu_5^2 - \frac{(\eta + m_c)^2}{2} \bigg] \ln \bigg[ \frac{(\sqrt{\mu^2 - (\eta + m_c)^2} + \mu)^2}{(\eta + m_c)^2} \bigg]\Bigg\},
\end{aligned}
\end{equation}
and
\begin{eqnarray}
    I_3(&&T = 0, \mu, \mu_5) = \frac{\theta(\mu - \eta - m_c)}{3\pi^2} \sqrt{\mu^2 - (\eta + m_c)^2} \nonumber\\
    &&\times \Big[ \mu^2 - (\eta + m_c)^2 + 3\mu_5^2 \Big].
\end{eqnarray}

\subsection{The optimal variational parameters $\overline \eta$ and $\overline \zeta$}

Under the assumptions made above, the variational criterion involves two parameters with canonical dimension of energy, $\eta$ and $\zeta$. Once the free energy density $\mathcal{F}$ is computed up to a given order $k$ in the OPT expansion, the remaining dependencies on $\eta$ and $\zeta$ are fixed through an optimization prescription, such as the principle of minimal sensitivity (PMS)~\cite{Stevenson:1981vj,Stevenson:1982wn,Kneur:2007vj,Kneur:2007vm},
\begin{equation}
\frac{d\mathcal{F}_{\text{OPT}}^{(k)}}{d\eta}\Biggl|_{\delta = 1} =0 \;\;\;\;\;\;\;{\rm and}\;\;\;\;\;\; \frac{d\mathcal{F}_{\text{OPT}}^{(k)}}{d\zeta}\Biggl|_{\delta = 1} = 0\,,
\label{pms1}
\end{equation}
which are adopted here.
Beyond its inherent simplicity and the possibility to yield results beyond the mean field, a key advantage of employing the $\mathcal{O}(\delta)$ approximation lies in the analytical derivation of the equations for the optimal parameters $\overline \eta$ and $\overline \zeta$ in terms of the previously defined integrals $I_2$ and $I_3$, which gives\footnote{We have neglected solutions that are independent of the coupling since these may be considered nonphysical, apart from not reproducing the LN result when $N_c \to \infty$  (see Ref.~\cite {Kneur:2010yv} for details).}
\begin{equation}
\begin{aligned}
    \Bigl[ \eta - (M - m_c) - G(\eta + m_c) I_2(T, \tilde\mu, \mu_5) \Bigr]\Bigr|_{\eta = \bar\eta, \zeta = \bar\zeta} = 0,
\label{etabar}
\end{aligned}
\end{equation}
and
\begin{equation}
\begin{aligned}
    \Bigl[ \zeta + 2G I_3(T, \tilde\mu, \mu_5) \Bigr]\Bigr|_{\eta = \bar\eta, \zeta=\bar\zeta} = 0.
    \label{zetabar}
\end{aligned}
\end{equation}
Notice that in a typical LN evaluation, one considers $G$ to be of order $N_c^{-1}$, such that in the limit $N_c \to \infty$, one obtains $\overline \eta = M-m_c$ and $\overline \zeta =0$. By substituting these optimal values into Eq.~(\ref {freeOPT}), one exactly retrieves the standard LN result, which is a reassuring result as far as the optimization procedure is concerned.

{}Let us recall that the constituent quark mass $M$, in each method, is obtained by minimizing the corresponding thermodynamic potential, $\frac{\partial \mathcal F}{\partial M} = 0$, resulting in the gap equation,
\begin{equation}
    M = m_c + 2G N_f N_c \mathcal{M} I_2(T, \tilde\mu, \mu_5),
    \label{gap}
\end{equation}
where, for convenience, we define $\mathcal{M} = {\overline\eta} + m_c$. The corresponding equation for LN is obtained by replacing $\mathcal{M}\to M$ and ${\tilde \mu} \to \mu $. Then, using Eq. (\ref{gap}) together with Eq. (\ref{etabar}) allows us to write the following self-consistent equation for the OPT effective mass 
\begin{equation}
    {\cal M}= m_c +2 G N_f N_c {\cal M} \left ( 1 + \frac{1}{2N_f N_c} \right ) I_2(T, \tilde\mu, \mu_5) \,,
\end{equation}
where  ${\tilde \mu} = \mu + {\overline \zeta}$ can be written as a second self-consistent relation. In this case, the   OPT effective chemical potential reads 
\begin{equation}
    {\tilde \mu } = \mu - 2 G I_3(T, \tilde\mu, \mu_5) \,.
    \label{mueff}
\end{equation}
Notice that by replacing $G \to N_c N_f G_V$ in Eq. (\ref{mueff}), one gets an effective chemical potential that is very similar to the one considered in mean field evaluations of the NJL model extended with a repulsive vector channel \cite {Buballa:2003qv}. Once again, this demonstrates that even if one considers $G_V=0$, as we do here,  the presence of exchange terms can radiatively generate vector contributions that are $N_c^{-1}$ suppressed \cite {Klevansky:1992qe}.
{}Finally, concerning the OPT free energy, let us point out that  $I_2$ relates to the scalar condensate $\langle {\overline \psi} \psi \rangle$, while $I_3$  relates to the vector condensate $n=\langle {\psi}^+ \psi \rangle$. Therefore, the OPT and LN free energies display physically distinct structures since the latter depends solely on $\langle {\overline \psi} \psi \rangle$  \cite {Ferroni:2010ct,Restrepo:2014fna}.  

\subsection{Regularization procedure}
\label{regularization}

A critical aspect of nonrenormalizable models concerns the regularization procedure that is typically implemented by introducing a sharp cutoff $\Lambda$.  This UV regulator is then considered to represent a parameter whose numerical value can be phenomenologically fitted from physical observables. It defines the upper energy scale at which most model predictions can be trusted.  
The naive introduction of a UV cutoff in the divergent momentum integrals
$I_1(T,\tilde{\mu}, \mu_5)$ and $I_2(T,\tilde{\mu}, \mu_5)$ like 
\begin{equation}
    \int \frac{d^3p}{(2\pi)^3} \frac{1}{E_p} \to\int^\Lambda \frac{d^3p}{(2\pi)^3} \frac{1}{E_p},
\end{equation}
is referred to as the TRS, with the superscript $\Lambda$ denoting a three-dimensional momentum cutoff. Here, we compare these results with the MSS approach, which applies the regularization exclusively to the vacuum terms. This procedure, first applied in color superconductivity studies~\cite{Farias:2005cr} and widely used in the description of the QCD phase diagram in different contexts~\cite{Farias:2016let,Duarte:2018kfd,Avancini:2019ego,Lopes:2021tro,Azeredo:2024sqc,Lopes:2025rvn}, ensures that vacuum and medium effects (coming from terms with dependencies on $T,\mu,\mu_5$, etc.) are properly separated from the divergent integrals prior to regularization. 

{}For example, within the MSS, the divergent integral $I_2(T,\tilde{\mu}, \mu_5)$ appearing in the gap and PMS equations is replaced by (see \cite{Farias:2016let} for further details)
\begin{eqnarray}
    \sum_{s = \pm 1} \int &&\frac{d^3p}{(2\pi)^3} \frac{1}{E_p} = 4 \bigg\{  I_\text{quad}(\mathcal{M}_0) \nonumber\\
    &&+ \Big[ 2\mu_5^2 - (\eta + m_c)^2 + \mathcal{M}_0^2 \Big] I_\text{log}(\mathcal{M}_0) \nonumber\\
    &&- \frac{2\mu_5^2 + (\eta + m_c)^2 - \mathcal{M}_0^2}{16\pi^2} \nonumber\\
    &&+ \frac{(\eta + m_c)^2 - 2\mu_5^2}{16\pi^2} \log \bigg( \frac{(\eta + m_c)^2}{\mathcal{M}_0^2} \bigg) \bigg\},
\label{I2div}
\end{eqnarray}
with $\mathcal{M}_0 = \mathcal{M}(T=0,\mu=0,\mu_5=0)$ being the vacuum mass (in the LN limit, $\mathcal{M}_0 \rightarrow M_0$). 
In Eq.~(\ref{I2div}), the functions $I_{\text{quad}}(M_0)$ and $I_{\text{log}}(M_0)$ are given in terms of divergent integrals that are expressed solely in terms of vacuum quantities, namely $\mathcal{M}_0$,
\begin{eqnarray}
    &&I_{\text{quad}}(\mathcal{M}_0) 
    = \frac{1}{2}\int^{\Lambda} \frac{d^3p}{(2\pi)^3} \frac{1}{\sqrt{|\textbf{p}|^2+\mathcal{M}^2_0}},\\
    &&I_{\text{log}}(\mathcal{M}_0) = - \frac{\partial}{\partial \mathcal{M}_0^2} I_{\text{quad}}(\mathcal{M}_0).
\end{eqnarray}
Consequently, $I_1$ can then be expressed as
\begin{eqnarray}
    \sum_{s = \pm 1}\int &&\frac{d^3p}{(2\pi)^3} E_p = 2 (\eta + m_c)^2 \Bigg\{ I_\text{quad}(\mathcal{M}_0) \nonumber\\
    &&+ \bigg[2\mu_5^2 - \frac{(\eta + m_c)^2}{2} + \mathcal{M}_0^2 \bigg] I_\text{log}(\mathcal{M}_0) \nonumber\\
    &&- \frac{3(\eta + m_c)^2 - 4\mathcal{M}_0^2 }{64\pi^2} \nonumber\\
    && + \frac{(\eta + m_c)^2 - 4\mu_5^2}{32\pi^2} \log\bigg[ \frac{(\eta + m_c)^2}{\mathcal{M}_0^2} \bigg]\Bigg\}.
\end{eqnarray}

The model parameters, represented by the current quark mass $m_c$, the scalar coupling constant $G$, and the three-dimensional cutoff $\Lambda$, are generally fixed to reproduce the empirical values of the pion decay constant $f_{\pi}$, the pion mass $m_{\pi}$ and the quark condensate $\langle\bar{q} q\rangle$. In this work, we take $\Lambda$ as input and determine the values of $m_c$ and $G$ that reproduce $f_{\pi} = 92.4$ MeV and $m_{\pi} = 135$ MeV. The parameter values considered in our numerical analysis are presented in Table ~\ref{parameter_table}.

\begin{table}[h]
\caption{Parameter sets for the LN and OPT approximations considered in this work. The parameters $G$ and $m_c$ are fitted using $\Lambda$ as an input to reproduce $m_\pi = 135 \; \text{MeV}$ and $f_\pi = 94.2 \; \text{MeV}$. The mass parameters, $\Lambda$ and $- \langle\bar{q} q\rangle^{1/3}$ are given in units of MeV.}
\label{parameter_table}
\begin{tabular*}{\columnwidth}{@{\extracolsep{\fill}}lccccc@{}}
\hline \hline
Model & $\Lambda$ (input) & $G \Lambda^2$ & $m_c$ & $M_q$ & $-\langle \bar{q} q\rangle^{1/3}$ \\ 
\hline
OPT & 635.0 & 1.99 & 5.1 & 299.8 & 246.2 \\
LN  & 640.0 & 2.14 & 5.2 & 321.1 & 247.2 \\
\hline \hline
\end{tabular*}%
\end{table}

\section{Numerical results}
\label{NR}

In this Section, we present the numerical results for the order parameters, thermodynamic quantities, and phase diagram, comparing the LN and OPT approaches. 

\begin{figure}
\centering
\includegraphics[width=0.95\linewidth]{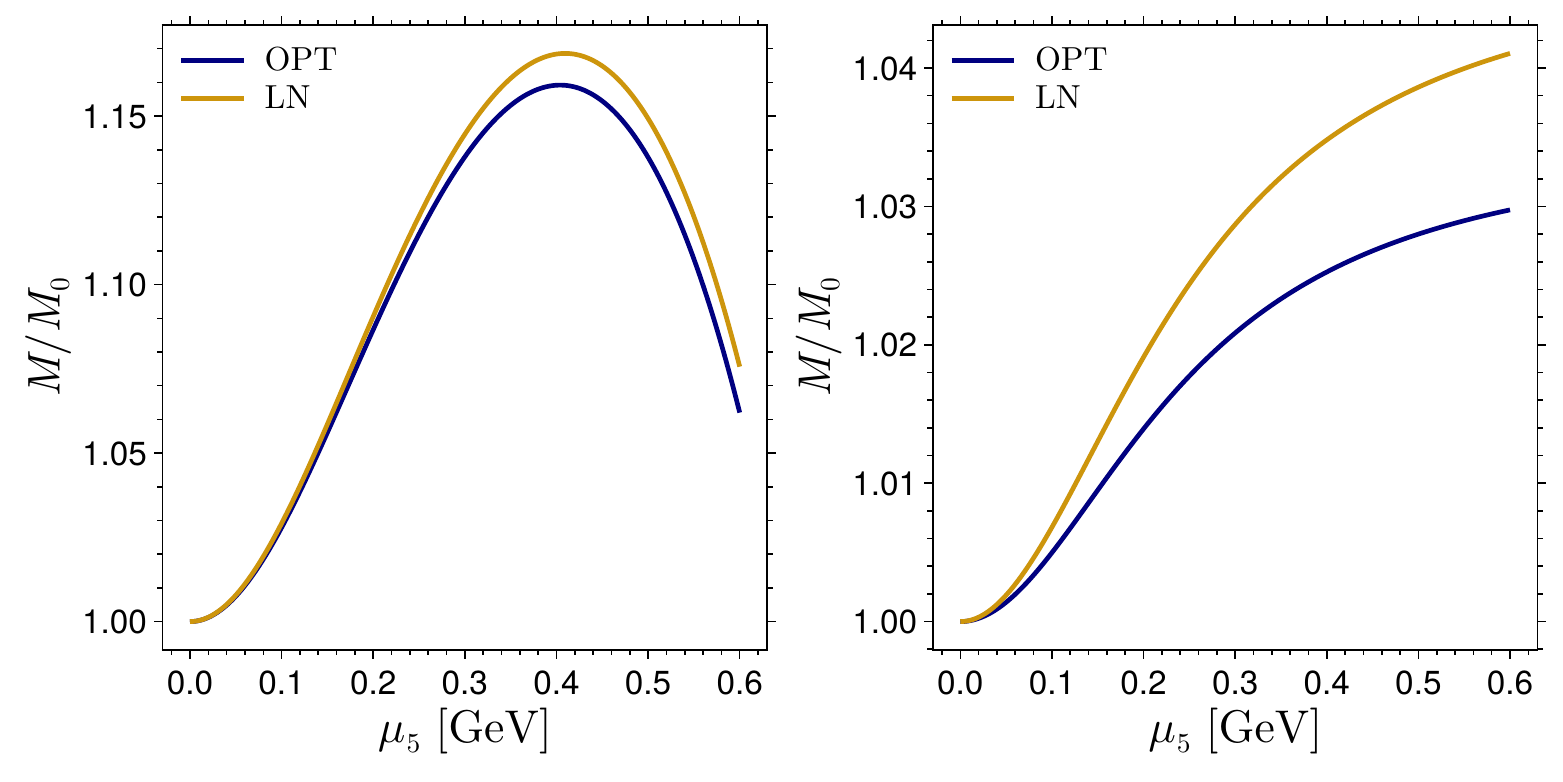}
    \caption{Effective quark mass $M$ normalized by its value in the vacuum, $M_0$, as a function of the chiral chemical potential $\mu_5$ at $T = \mu = 0$. The left panel shows the results for TRS, while the right panel shows the MSS predictions. In both panels, dark and light curves correspond to OPT and LN results, respectively.}
\label{fig1}
\end{figure}

It is instructive to begin the analysis with the zero baryon density case.
The left panel of {}Fig.~\ref{fig1} shows the results for the effective quark mass $M$ as a function of the chiral chemical potential $\mu_5$ at zero temperature for both LN and OPT cases when considering the TRS regularization procedure, while the right panel shows the results when considering the MSS case. The normalization $M_0$ is the value of $M = \bar{\sigma} + m_c$ at $T = \mu = \mu_5 = 0$, (and also the scale for MSS). One can observe that in both LN and OPT approaches, the use of TRS leads to an initial increase in $M$ with $\mu_5$, followed by a decrease at $\mu_5 \gtrsim  0.42 \,{\rm GeV}$. In contrast, when the MSS is employed, both methods exhibit a monotonic increase in the effective quark mass, which is a behavior that is in accordance with previous studies comparing TRS and MSS within the mean field approach~\cite{Farias:2016let}. 


\begin{figure}
    \centering
\includegraphics[width=0.95\linewidth]{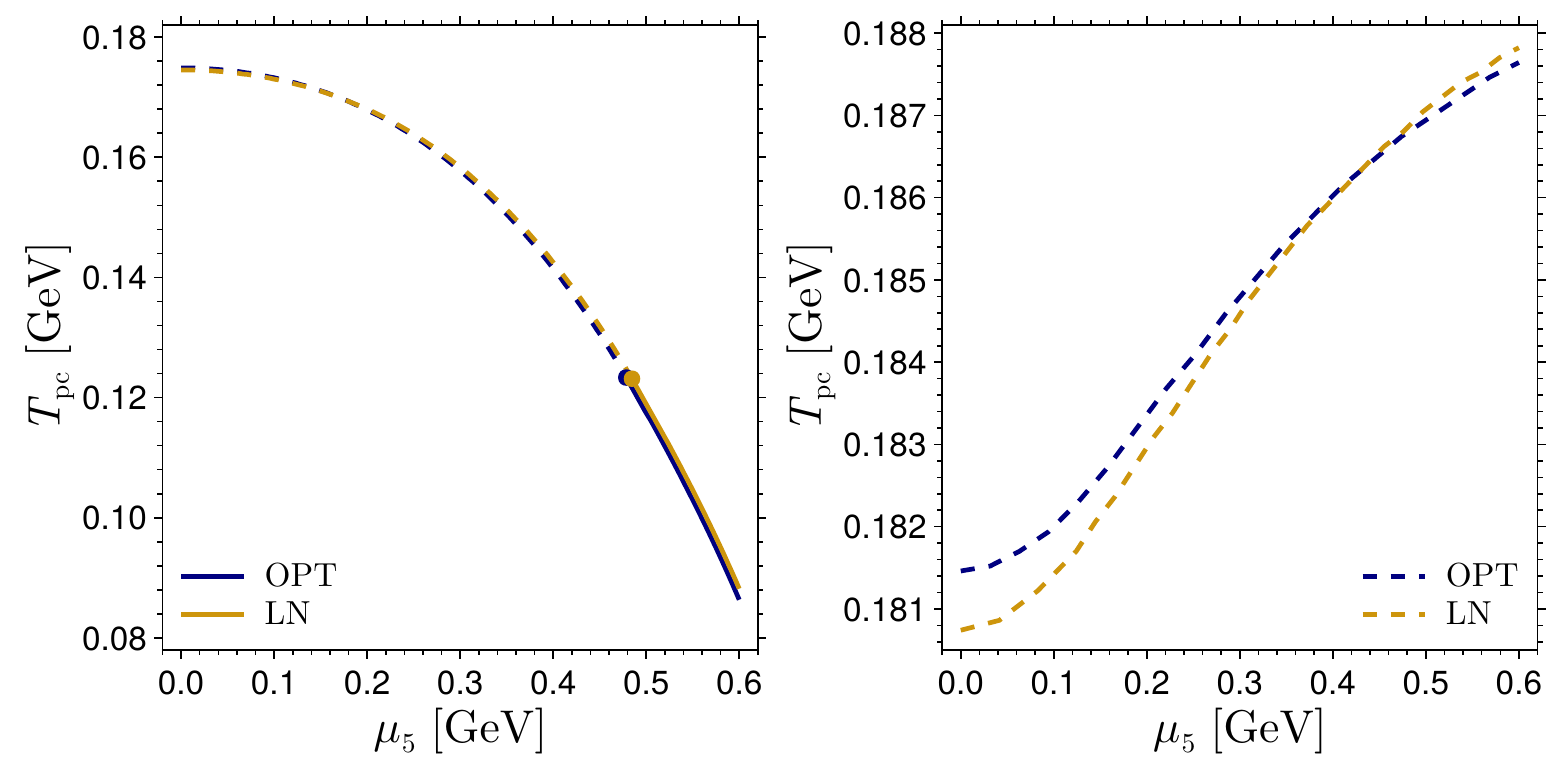}
    \caption{Phase diagrams on the $T_\text{pc} \times \mu_5$ plane at $\mu = 0$ comparing LN (light) and OPT (dark). The left panel shows the results for TRS, while the right panel shows the MSS predictions. The continuous lines represent first-order transitions terminating at a critical endpoint, while dashed lines represent crossover transitions.}
    \label{fig2}
\end{figure}

The phase diagrams in the pseudocritical temperature\footnote{Recall that when considering the case of finite quark masses ($m_c \ne 0$), as we do here, one observes a crossover taking place at a pseudocritical temperature, $T_{pc}$.}  ($T_\text{pc}$)  versus $\mu_5$ plane are shown in both panels of {}Fig~\ref{fig2} for both the LN and OPT approximations. The results obtained with MSS (right panel) are consistent with lattice QCD simulations at finite $\mu_5$~\cite{Braguta:2015owi}, as well as with theoretical model predictions that properly implement the regularization of divergent integrals~\cite{Farias:2016let,Azeredo:2024sqc}. In this case, there is no critical endpoint, and the pseudocritical temperature for chiral symmetry restoration increases monotonically with $\mu_5$, in contrast to the TRS result (left panel), which exhibits the opposite qualitative behavior. 
{}For completeness, let us point out that there is almost no difference in the $T_{pc}\times\mu_5$ diagram for the TRS with OPT and LN; while within the MSS scheme, the pseudocritical temperature values for OPT are higher in the range $\mu_5\lesssim 0.35$ GeV compared to the LN values. 


\begin{figure}
\centering
\includegraphics[width=0.95\linewidth]{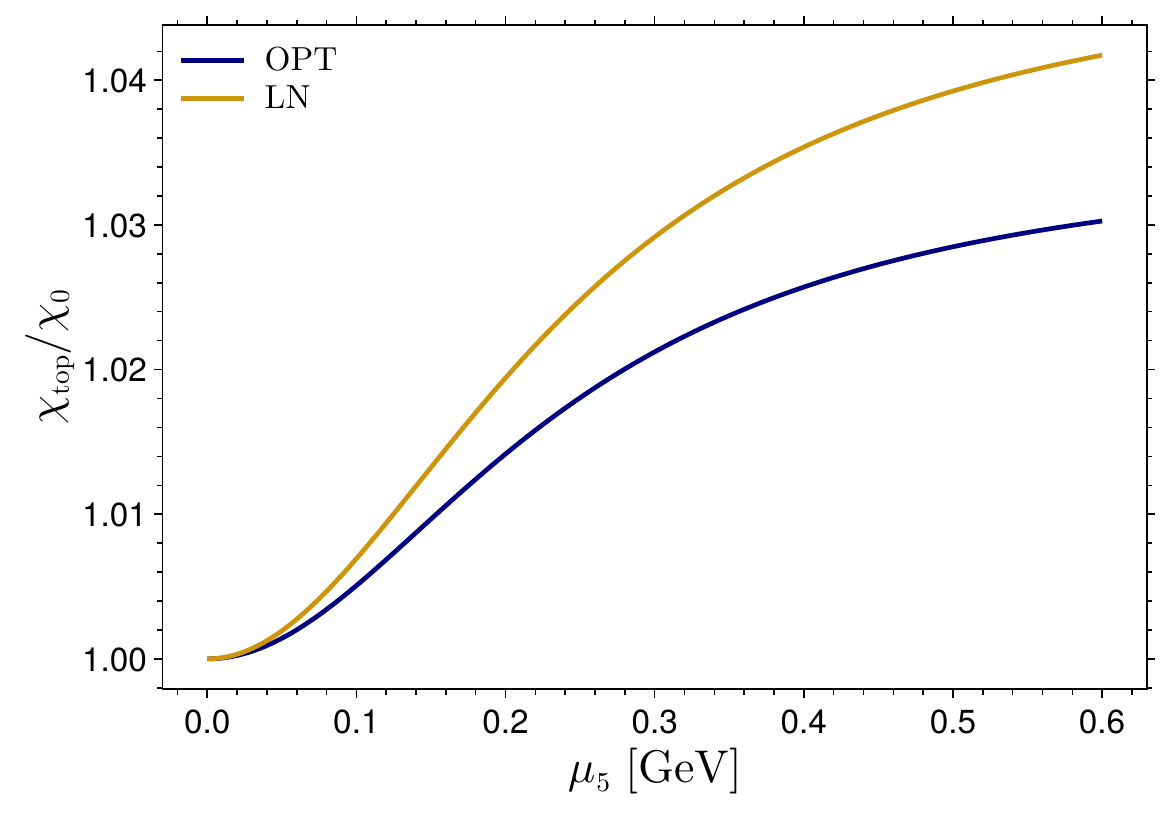}
 \caption{Normalized topological susceptibility $\chi_\text{top}/\chi_0$ as a function of the chiral chemical potential $\mu_5$ at zero temperature and baryon density, comparing LN (light) and OPT (dark) within the MSS framework. Here $\chi_0$ is the value of $\chi_\text{top}$ at $T = \mu = \mu_5 = 0$.}
    \label{fig3}
\end{figure}

In the presence of a chiral imbalance, which may arise due to topological fluctuations in the gluonic fields, the topological susceptibility $\chi_{\text{top}}$ becomes a relevant quantity for analysis. 
In fact, it has been observed that the increase of $\mu_5$ leads to a suppression of $\chi_{\text{top}}$ at finite temperatures, as a consequence of the fact that a chiral imbalance disfavors topological configurations such as instantons, which are responsible for the axial $U(1)$ anomaly~\cite{Ruggieri:2020qtq}. At zero temperature, the topological susceptibility is defined as
\begin{equation}
    \chi_\text{top} = m_c |\left< \bar\psi \psi \right>|.
    \label{chitop}
\end{equation}  
Since $\chi_{\text{top}}$ is proportional to the absolute value of the chiral condensate, it is expected to increase with $\mu_5$. 
{}Figure ~\ref{fig3} illustrates this behavior for both LN and OPT within the MSS approach. This qualitative behavior is consistent with previous results from effective models~\cite{Ruggieri:2020qtq} and lattice QCD simulations~\cite{Braguta:2015owi,Astrakhantsev:2019wnp}. Therefore, the expected behavior for the topological susceptibility is obtained here 
only by the MSS prescription, whereas the TRS
does not give the correct result.


\subsection{Results at finite quark density and zero temperature}

Here, we extend our analysis to finite densities. Since the role of MSS is to obtain the correct behavior of the thermodynamic quantities, we restrict ourselves to presenting only the results for MSS, performing comparisons between the LN and OPT approaches. 


\begin{figure}
    \centering
    \includegraphics[width=0.95\linewidth]{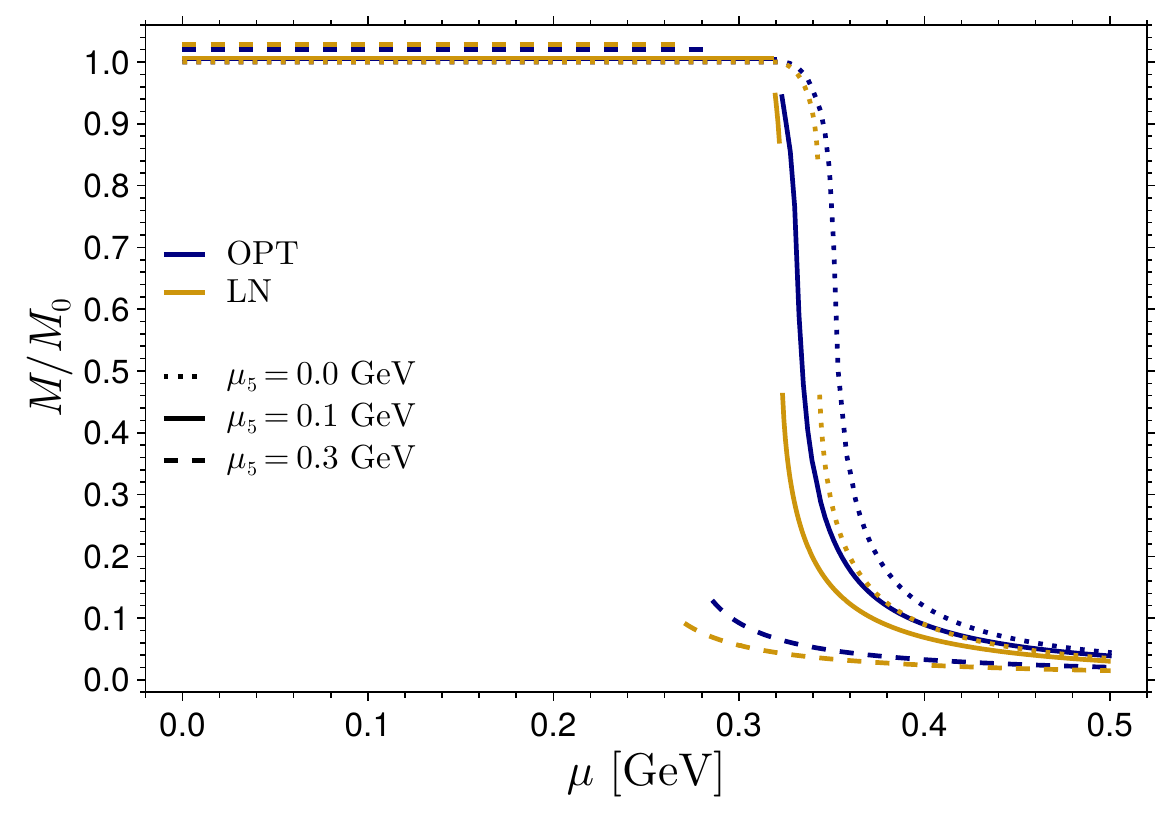}
    \caption{Effective quark mass $M$ normalized by its in vacuum value, $M_0$, as a function of quark chemical potential for both LN (light) and OPT (dark).}
    \label{fig4}
\end{figure}

{}Figure~\ref{fig4} shows the behavior for the effective quark mass $M$ normalized by its in-vacuum value $M_0$  as a function of the quark chemical potential at zero temperature and for different values of $\mu_5$. 
{}For zero chiral chemical potential, the chiral phase transition exhibits different behaviors depending on the approximation scheme: within the LN approach, there is a weak first-order transition, whereas in the OPT framework, a smooth crossover is observed in accordance with our previous discussion. The smoothening of the transition in the OPT approximation is also a result of applying the MSS that, even in the zero $\mu_5$ case, works to separate implicit medium dependencies present through the dependence on $M = M(T, \mu, \mu_5)$.

At nonzero values of $\mu_5$, a first-order phase transition associated with partial restoration of chiral symmetry is observed in all the cases considered in this work, with the discontinuity becoming more pronounced for higher values of $\mu_5$. This enhancement of the first-order nature is more evident within the LN approximation, which tends to produce stronger phase transitions when compared to the OPT. {}Furthermore, it may be observed that the chiral chemical potential for the chiral transition decreases as $\mu_5$ increases. 
At a finite chiral chemical potential, e.g., $\mu_5\lesssim 0.1$ GeV, the LN approximation shows a second discontinuity, which is not found in the OPT method. It is relevant to note that the additional discontinuity in the LN case tends to merge with the first one as we increase the value of the chiral chemical potential.

\begin{figure}
    \centering
    \includegraphics[width=0.95\linewidth]{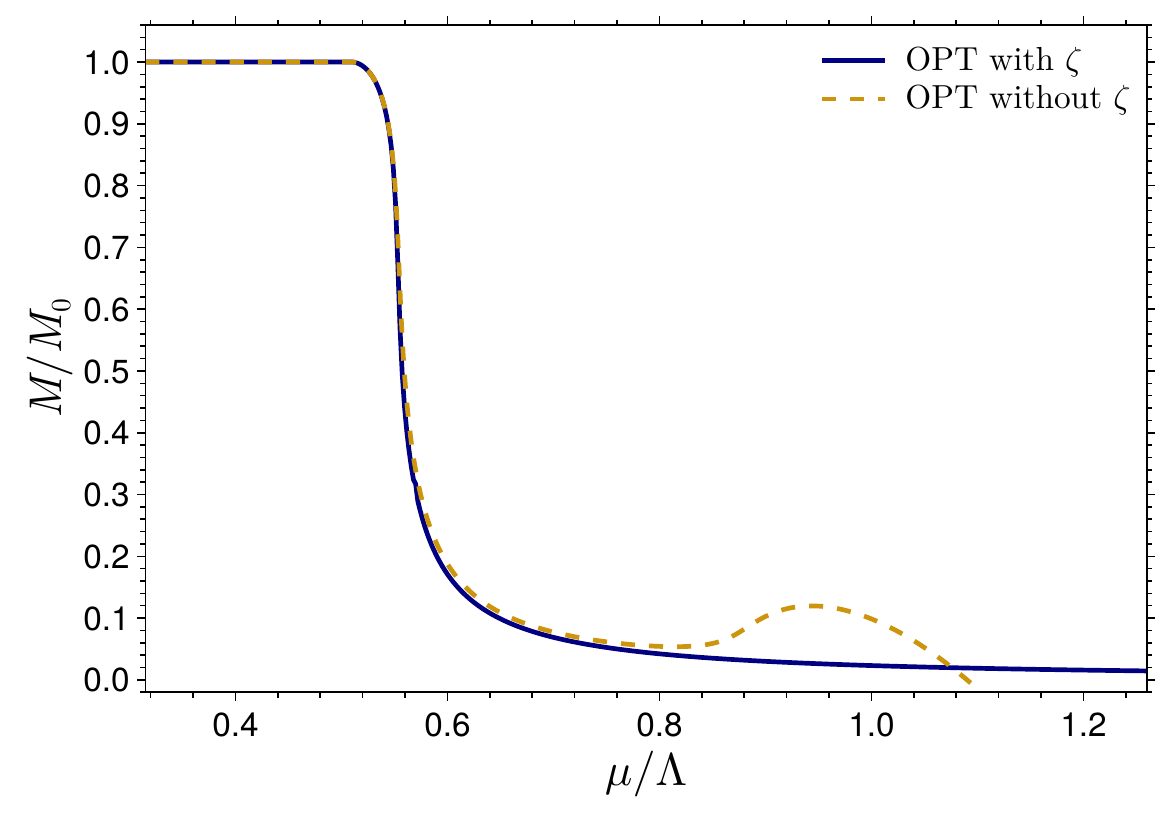}
    \caption{Effective quark mass $M$ normalized by the vacuum mass $M_0$, as a function of  $\mu/\Lambda$. The inclusion of $\zeta$ assures that the OPT effective mass behaves well as one approaches the highest possible density values allowed by the model ($\mu \to \Lambda= 0.635$ GeV).}
    \label{zeta_effect}
\end{figure}

At this point, it is important to illustrate how the additional variational parameter, $\zeta$, preserves the stability of the optimization process at high densities ($\mu \to \Lambda$). The result shown in {}Fig.~\ref {zeta_effect} indicates that by following the procedure originally outlined in Ref. ~\cite{Kneur:2010yv}, where $\zeta=0$, and employing the parameter set discussed in Sec.~\ref{regularization}, one observes that around $\mu \approx 0.5$ GeV, the dynamical quark mass begins to increase with $\mu$. Then, it reaches a maximum value at $\mu \approx \Lambda$ before diving into a region where $M<0$. This behavior is clearly nonphysical, as an increasing density is expected to drive chiral symmetry restoration. The result shown in {}Fig.~\ref {zeta_effect} illustrates well how this problem is effectively solved by including the second variational parameter
$\zeta$, highlighting that the replacement $\mu\to \mu+ (1-\delta)\zeta$ in the partition function is crucial to accurately describe the high-density regime within the OPT framework.

{}From the free energy density, given by Eq.~(\ref{freeOPT}), all the thermodynamic quantities follow. The pressure density is $P= - {\cal F}(T,\mu_B,\mu_5)$ and the energy density $\epsilon$ is
\begin{equation}
    \epsilon = -P + Ts + \mu_B n_B + \mu_5 n_{5},
\label{epsilon}
\end{equation}
where the entropy, baryon and chiral densities are defined, respectively, as\footnote{Note that, as the OPT free energy density Eq.~(\ref{freeOPT}) also depends on the variational parameters $\eta$ and $\zeta$, the defining expressions for
$s,\; n_B$, and $n_5$ explicitly also involve partial derivatives of ${\cal F}_{\rm OPT}$ with respect to $\eta$ and $\zeta$. However, thermodynamic consistency is nonetheless ensured since the PMS conditions (\ref{pms1}) automatically cancel these contributions. The expressions for $s,\; n_B$, and $n_5$ then follow from the standard thermodynamic relations. }
\begin{eqnarray}
&&s= - \frac{\partial {\cal F}}{\partial T},
\label{st}
\\
&&n_B= \frac{\partial P}{\partial \mu_B},
\label{nB}
\\
&&n_5= \frac{\partial P}{\partial \mu_5}.
\label{n5}
\end{eqnarray}
Note that the baryonic density is related to the density $n$ by $n_B=n/3$. We also recall here that the chiral density is divergent. Such divergence arises from zero-point fermionic fluctuations, which acquire explicit dependence on the chiral chemical potential $\mu_5$. Although this contribution vanishes for massless fermions, it becomes logarithmically divergent for massive ones, reflecting the thermodynamically incompatibility between a finite value of the chiral density and the absence of the chiral symmetry for massive fermions~\cite{Ruggieri:2020qtq}.
Therefore, in the presentation of all our 
numerical results, we have subtracted this divergent term.

\begin{figure}
    \centering
    \includegraphics[width=0.95\linewidth]{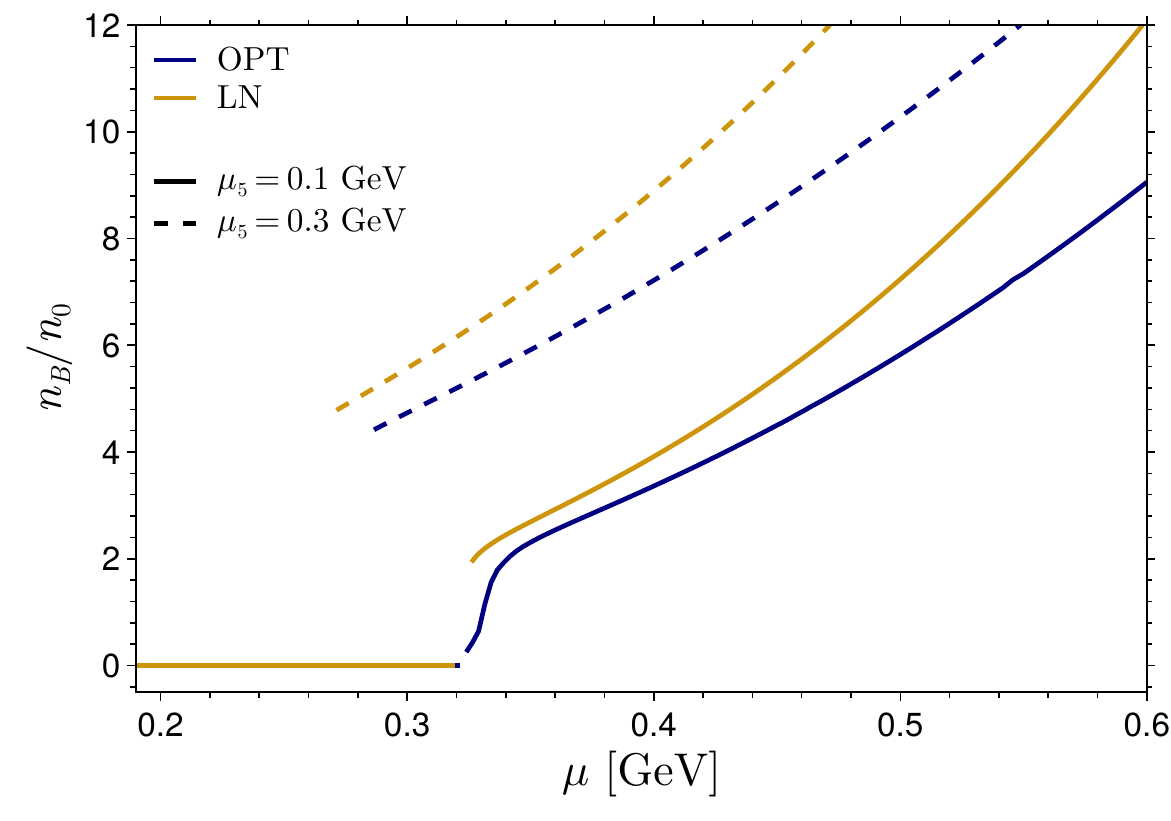}
    \caption{Baryonic density $n_B$, normalized by the nuclear saturation density $n_0 = 0.17 \; \text{fm}^{-3}$, as a function of the quark chemical potential, $\mu$, comparing the LN (light lines) and OPT (dark lines) approaches.}
    \label{fig6}
\end{figure}

In {}Fig.~\ref{fig6}, we show the baryon density $n_B$ normalized by the nuclear density $n_0 = 0.17 \; {\rm fm}^{-3}$ and as a function of $\mu$. We can also observe here
the features of the first-order transition discussed previously. The strongest discontinuity is observed for $\mu_5 = 0.3$ GeV, where the baryon density jumps from 0 to $\gtrsim 4.5 n_0$ in both approximations, while the jump is more pronounced in the LN than in the OPT case for $\mu_5 = 0.1$ GeV. 


\begin{figure}
    \centering
\includegraphics[width=0.95\linewidth]{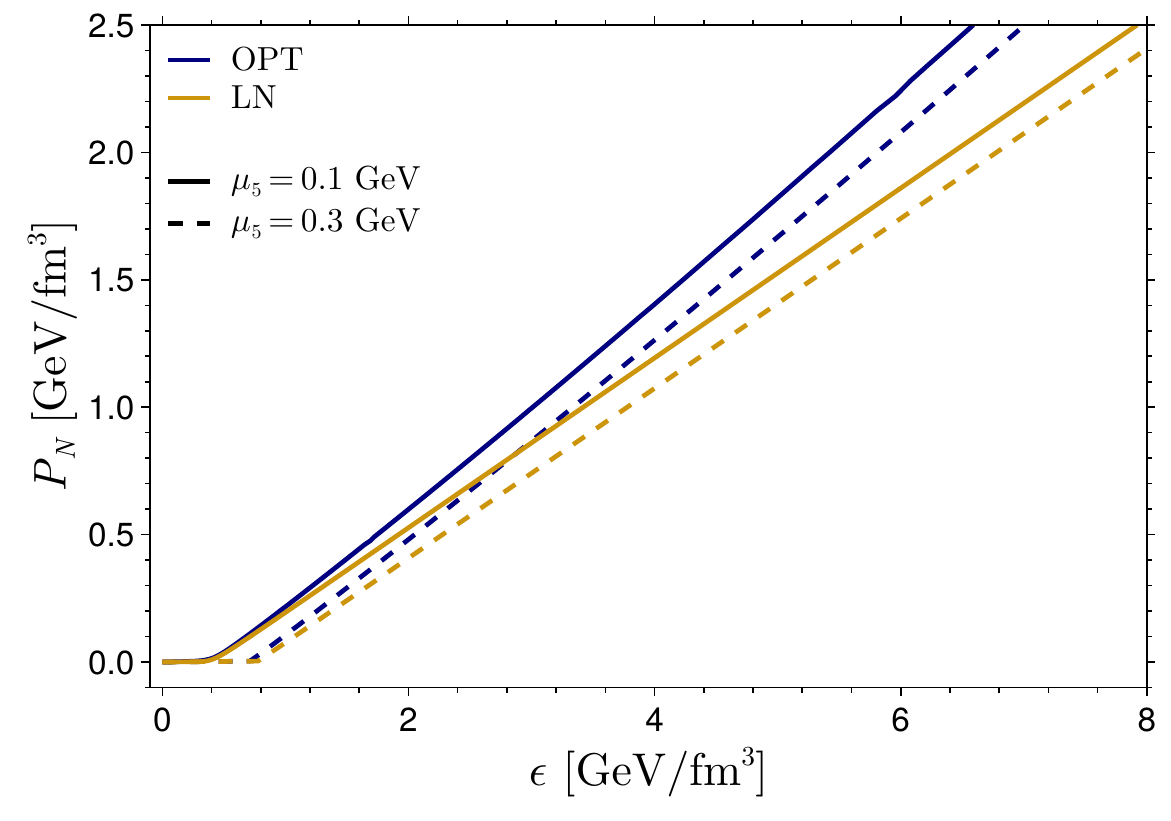}
    \caption{Equation of state, for both LN (light line) and OPT (dark line).}
    \label{fig7}
\end{figure}

The results for the equation of state $P\times\epsilon$ (EoS) are shown in {}Fig.~\ref{fig7}. In this figure we have defined $P_N = - [{\cal F}(T,\mu_B,\mu_5) - {\cal F}(0,0,\mu_5)]$ so that both  pressure and energy density are zero when $\mu = 0$.
{}As anticipated,  for both $\mu_5$ values considered, the respective EoS is noticeably stiffer in the OPT approximation compared to the LN case, throughout the entire range of $\mu$ values considered.  It should also be noted that within a given approximation scheme, the increase of $\mu_5$ leads to a softening of the EoS. Moreover, it is important to mention that the behavior observed for OPT in Ref.~\cite{Kneur:2010yv} is also attributed to the absence of the second variational parameter $\zeta$, as previously discussed at the beginning of this Section. In the absence of $\zeta$ the results of  Ref.~\cite{Kneur:2010yv} show that the OPT produces a very stiff EoS at low densities, followed by a peaklike structure, after which the stiffness was notably moderated, rendering it softer than the LN curve for $\mu \gtrsim \Lambda$. On the other hand, with the replacement $\mu\to \mu+ (1-\delta)\zeta$, the OPT framework now yields a stiffer EoS across all investigated density regimes.


\begin{figure}
    \centering
\includegraphics[width=0.95\linewidth]{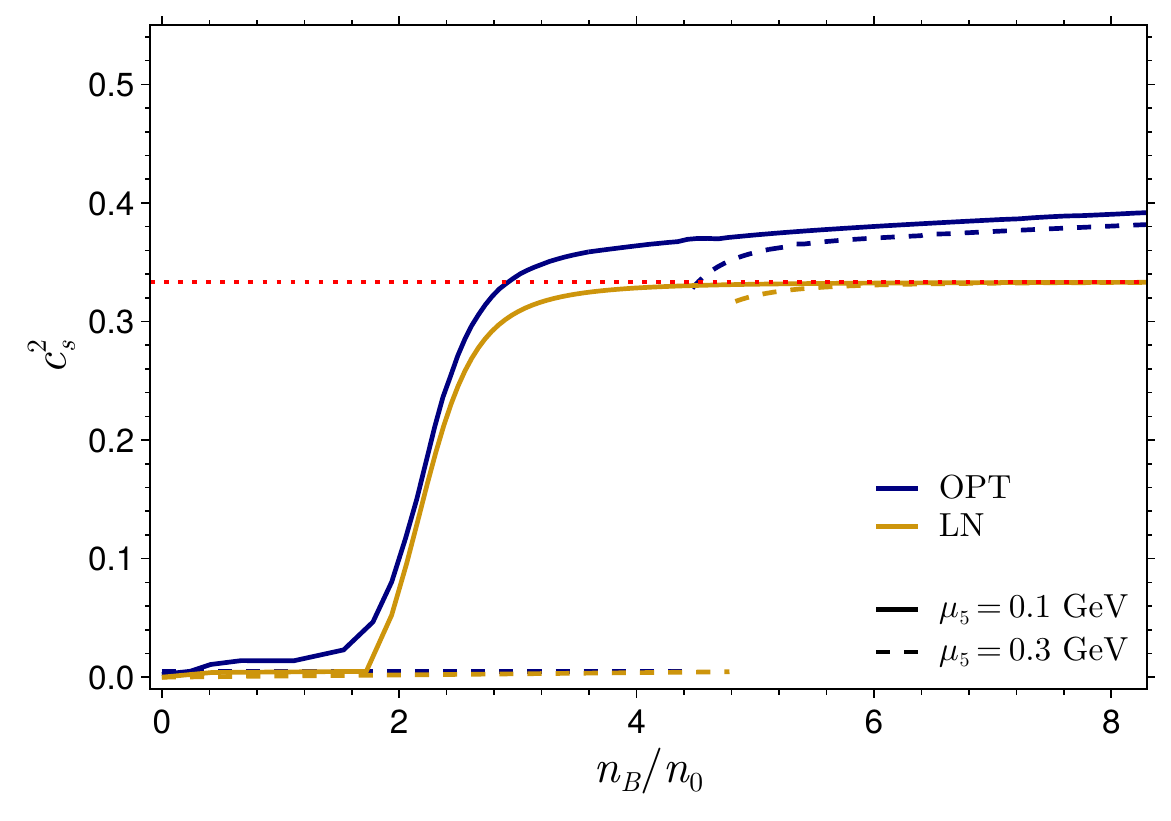}
    \caption{Speed of sound squared, as a function of baryonic density, for different values of $\mu_5$. Dark and light curves correspond to OPT and LN results, respectively.}
    \label{fig8}
\end{figure}

\begin{figure}
    \centering
\includegraphics[width=0.95\linewidth]{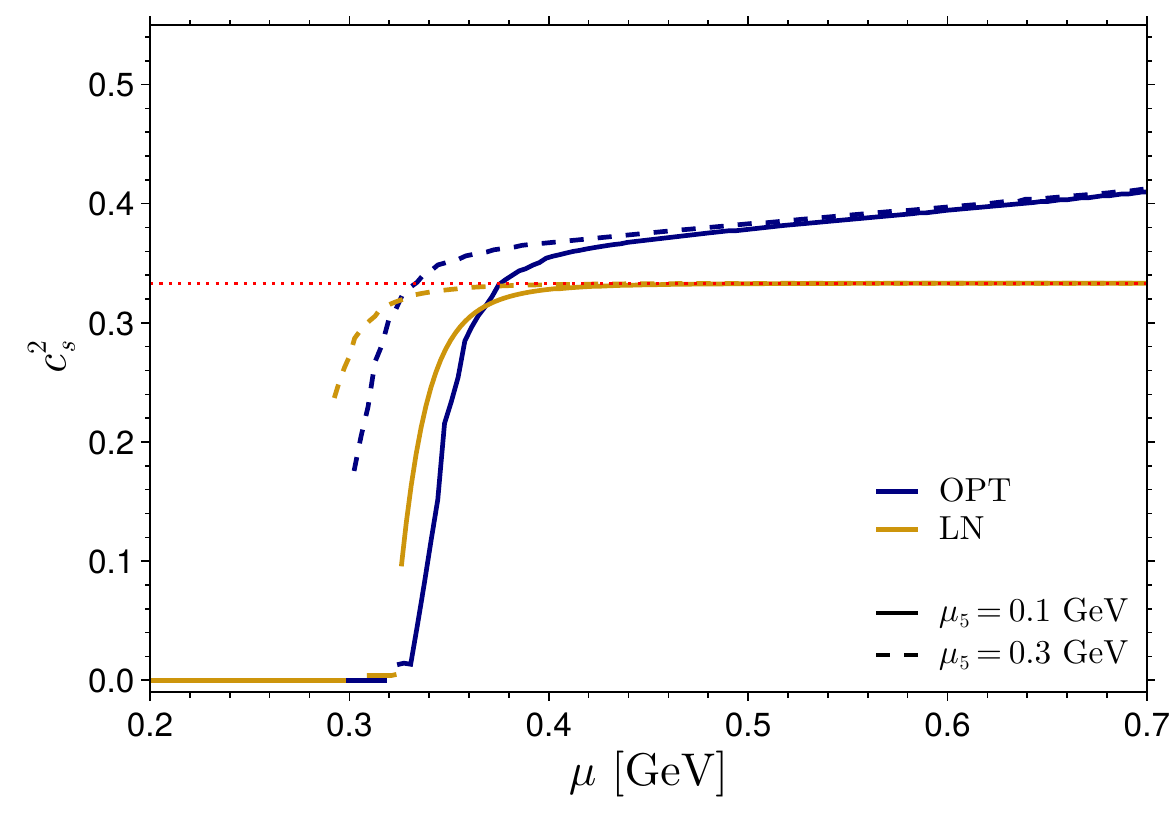}
    \caption{Speed of sound squared, as a function of quark chemical potential, for different values of $\mu_5$. Dark and light curves correspond to OPT and LN results, respectively.}
    \label{fig9}
\end{figure}

In {}Figs.~\ref{fig8} and \ref{fig9} we show the results for the speed of sound squared, 
\begin{equation}
c_s^2 = \frac{\partial P_N}{\partial\epsilon},
\end{equation}
as a function of the normalized baryon density and the chemical potential, respectively. {}For $\mu_5 = 0.3$ GeV, it is observed that in the transition region, $c_s^2$ jumps from zero toward a value close to the conformal limit of 1/3. In contrast, for $\mu_5 = 0.1$ GeV, the first-order transition is weaker, resulting in a smaller jump in $c_s^2$. The increased stiffness of the EoS within the OPT framework, compared to the LN approximation, is also evident in the sound velocity.
{}For the two values of the chiral chemical potential considered, the sound speed exceeds the conformal value and does not appear to converge at asymptotically large densities. In contrast, the LN converges rapidly to 1/3 immediately after the transition. Although the stiffness of the EoS in the OPT at low-density regimes is adequate for reproducing the high masses inferred from gravitational wave observational data~\cite{LIGOScientific:2018cki,Miller:2019cac,Miller:2021qha}, the stiffness should ideally be moderated at higher densities to yield smaller radii, a behavior not observed in the current results. Achieving such smaller radii would require the inclusion of additional physical ingredients, which extend beyond the scope of this work.
For the reasons discussed in the previous section, the OPT generates a stiffer EoS, which, in turn, allows $c_s^2$ to reach values beyond the conformal value at high $\mu$ values, as the figure shows.


\section{Conclusions}\label{conclusions}

In this work, we have employed the MSS procedure to investigate the influence of a finite chiral chemical potential on the QCD phase structure at finite densities and temperatures within the OPT framework applied to the two-flavor NJL model. Our analysis focused on the behavior of thermodynamic quantities and order parameters, comparing results from the OPT and LN approximations.

At zero quark chemical potential, we revisited the basic phase structure of the pseudocritical temperature as a function of the chiral chemical potential, comparing the LN and OPT approximations. In the TRS prescription, the phase diagram is almost identical in both approximations, with the pseudocritical temperature decreasing as a function of the chiral chemical potential until a critical endpoint, which is followed by a first-order phase transition line. In contrast, in the MSS regularization, both approximations show an increase in the pseudocritical temperature as a function of the chiral chemical potential --- a result that is in qualitative agreement with well-established results from the LQCD and Dyson-Schwinger equations. In this case, the OPT approximation predicts higher values for the pseudocritical temperature in the range $\mu_5 \lesssim 0.35$ GeV compared to LN. The LN and OPT results that were obtained here using MSS regularization strongly disagree with the findings of Ref.~\cite{Ghosh:2023rft,Ghosh:2025zkk}, which predict the ICC effect, which can be attributed to a regularization artifact associated with TRS-type prescriptions. 

As expected, we have observed that the LN tends to predict stronger first-order phase transitions compared to the OPT for the range of $\mu_5$ values considered. For completeness, the physical reasons for such behavior have been reviewed following Refs.~\cite {Ferroni:2010ct,Restrepo:2014fna}. At the same time, we have found that the presence of a chiral chemical potential enhances the first-order nature of the chiral phase transition at zero temperature and lowers the quark chemical potential at which it occurs. {}Furthermore, we have shown that the inclusion of the additional optimization parameter $\zeta$ in the OPT proves to be essential to prevent nonphysical behavior at high densities.

Our results confirm that, also for chirally imbalanced quark matter, the presence of two-loop (exchange) contributions in the OPT free energy produce a stiff EoS, which may have implications for the study of dense matter in astrophysical contexts. Moreover, the observed sensitivity of the phase structure to model parameters reinforces the importance of a careful treatment of divergences in nonrenormalizable models.

{}Future investigations could include extensions to finite temperature and/or the incorporation of vector and diquark couplings, providing a more realistic description of extremely dense environments.

\begin{acknowledgments}

This work was partially supported by Conselho Nacional de
Desenvolvimento Cient\'{\i}fico e Tecnol\'ogico (CNPq), Grants
No. 312032/2023-4, No. 402963/2024-5 and No. 445182/2024-5 (R.L.S.F.), No. 307286/2021-5 (R.O.R.), No. 307261/2021-2  and No. 403016/2024-0 (M.B.P.); Funda\c{c}\~ao
de Amparo \`a Pesquisa do Estado do Rio Grande do Sul (FAPERGS),
Grants No. 24/2551-0001285-0 and No. 23/2551-0000791-6 (R.L.S.F.),
No. 23/2551-0000791-6 and No. 23/2551-0001591-9 (D.C.D.); Funda\c{c}\~ao
Carlos Chagas Filho de Amparo \`a Pesquisa do Estado do Rio de Janeiro
(FAPERJ), Grant No. E-26/201.150/2021 (R.O.R.), Grant No. SEI-260003/019544/2022
(W.R.T). The work is also part of the project
Instituto Nacional de Ci\^encia e Tecnologia - F\'isica Nuclear e
Aplica\c{c}\~oes (INCT - FNA), Grant No. 464898/2014-5 and supported
by the Ser\-ra\-pi\-lhei\-ra Institute (grant no. Serra -
2211-42230).

\end{acknowledgments}


\end{document}